\documentclass[10pt]{article}

\usepackage{amsmath}
\usepackage{amssymb, epsfig}

\usepackage{graphicx}

\usepackage{cite}

\usepackage{color}
\usepackage{soul}

\usepackage[labelfont=bf,labelsep=period,justification=raggedright]{caption}

\makeatletter
\renewcommand{\@biblabel}[1]{\quad#1.}
\makeatother

\date{}

\begin{document}

\begin{flushleft}
{\Large
\textbf{Stem-Like Adaptive Aneuploidy and Cancer Quasispecies}
}
\\
Domenico Napoletani$^{1,\ast}$,
Michele Signore$^{2}$,
Daniele C. Struppa$^{1}$
\\
\bf{1} 
Schmid College of Science and Technology, Chapman University, Orange, CA, 92866
\\
\bf{2} 
Department of Hematology, Oncology and Molecular Medicine, Tumor Stem Cell Biobank, Istituto Superiore di Sanit\`{a}, 00161, Rome, Italy
\\
$\ast$ E-mail: napoleta@chapman.edu
\end{flushleft}

\section*{Abstract}
We analyze and reinterpret experimental evidence from the literature to argue for an ability of tumor cells to self-regulate their aneuploidy rate. We conjecture that this ability is mediated by a diversification factor that exploits molecular mechanisms common to embryo stem cells and, to a
lesser extent, adult stem cells, that is eventually reactivated in tumor cells. Moreover,
we propose a direct use of the quasispecies model to cancer cells based on their significant
genomic instability (i.e. aneuploidy rate), by defining master sequences lengths as the sum of all copy numbers of physically distinct whole and fragmented chromosomes. We compute an approximate error threshold such that any aneuploidy rate larger than the threshold would lead to a loss of fitness of a tumor population, and we confirm that highly aneuploid cancer populations already function with aneuploidy rates close to the estimated threshold.

Keywords: Quasispecies, Cancer, Chromosomal instability, Aneuploidy.



\section*{Introduction}

Non-diploid chromosome content, also known as aneuploidy, is the most common feature of human tumor cells \cite{AneuCancer0, CancerCIN2, AneuCancer2}. However, there has been dispute on whether aneuploidy and chromosomal instability, i.e. the tendency to gain or lose parts of the genome during cell replication, give an advantage or disadvantage to the tumor cells, with recent evidence strongly siding the former \cite{AneuCancer3}.

Less attention has been given to potentially beneficial roles of aneuploidy in developmental biology, as it is generally assumed to be a byproduct of aberrant cell division, with mostly lethal or negative impact.
And yet, high levels of aneuploidy are associated to increased adaptability in plants and yeast \cite{PoliPloPlants,HeatShock}, and a certain rate of aneuploidy, leading to precise percentages of mosaic aneuploidy, is common in several mammals' embryos (\cite{MixoploAnimals} chapter 10), including humans \cite{ChaosEmbryo, HumanSC1}, much more common than the corresponding miscarriage rate would imply, at least if we extrapolate evidence from pigs embryos \cite{PigsEmbryos}. Similarly, it is speculated that the significant mosaic aneuploidy in adult human organs such as liver and brain is instrumental to an increased plasticity and adaptability of such organs \cite{liver,brain}, with \cite{AdultAndEmbryoMosaic} raising the possibility that the extensive aneuploidy in the embryo may transfer into similarly widespread copy number variations in all human tissues.

The specificity of aneuploidy in its manifestations is striking: a defined percentage of aneuploid cells, and a confinement of higher aneuploidy to specific organs rather than uniform distribution across the body. These facts are suggestive of a fine tuned, even post-embryonic, use of aneuploidy rather than a simple byproduct of aberrant or sustained cell division.

Since many types of cancers partially inherit the hierarchical structure of the tissues they have derived from and are assumed to be propagated by stem-like tumor cells \cite{CancerStem1}, it is possible that increased aneuploidy rates are used actively, to the population advantage, to increase the adaptability of stem or fast-dividing progenitor cells.

In this scenario a high chromosomal instability rate or gene copy-number variation, both resulting in mosaicism, would be the means to achieve enough genetic diversification, indeed we will refer to them as {\it aneuploidy rate} in this paper to emphasize even more their close link and we formally{\it define aneuploidy rate as the average probability that there is at least one new aneuploid modification per chromosome during cell replication}.

Observable levels of aneuploidy hold in adult cells \cite{AdultMosaic1}, even though much lower than in embryos. While this widespread aneuploidy could already originate during embryo development \cite{AdultAndEmbryoMosaic}, adult, non-transformed stem cells continue to have distinct levels of aneuploidy rates according to their type. For example, mesenchymal stem cells are likely to have very low aneuploidy rates \cite{AdultStemAneu}, while hepatocytes together with small intestine and pancreas cells display within-tissue extensive copy number variation (CNV) \cite{hepatocytesCIN, AdultTissueCNV}.

These strikingly different aneuploidy rates among embryo stem cells and adult stem cells raise the possibility that the finely tuned high aneuploidy rate observed in embryos is adaptively regulated by some mechanism specific to stem cells, that is accidentally reactivated in cancer cells, a {\it diversification factor}.

In Section 2 we broadly review existing literature on aneuploidy, collecting hints of the potentially positive impact of aneuploidy for complex organisms and cancer cells.  We then suggest specific evidence for the existence of a diversification factor by reinterpreting, in Section 3, recent single cell analysis experimental work.

The quasispecies model of evolution was introduced by Eigen in 1971 and has been applied in many different fields, on account of its usefulness as a general evolutionary model for error-prone self-replicative systems \cite{quasispeciesBiology}. Assuming aneuploidy rate is regulated in normal and cancer (stem) cells, quasispecies theory can be adapted to predict a maximal aneuploidy rate, an error threshold, after which each cancer subpopulation looses its identity, and therefore its ability to carry to future cells' generations its selectively advantageous genetic traits, what is referred to as error catastrophe \cite{quasispecies1}.

In Section 4 we estimate an error threshold for aneuploidy rates in cancer cells.Unlike previous attempts to adapt quasispecies theory to cancer \cite{QScancer1,QScancer2,QScancer3,QScancer4,QScancer5, QSNowak1, QSNowak2}, we do not recommend a specific alternative phenomenological model of cancer cells sub-populations dynamics. Instead,  we completely refocus the quasispecies model on aneuploidy, by defining the notion of chromosomal master sequences, whose length is taken to be the sum of the copy numbers of each whole or fragmented chromosome, and by using aneuploidy rates in the calculation of the probability of precise reproduction of sequences.

\section*{Aneuploidy in Normal and Cancer Stem Cells}

In normal cells the number of chromosomes and the total DNA content depends on the phase of the cell cycle \cite{MS28}. On the contrary, cancer cells usually display aneuploidy, and their chromosome load is generally higher than normal cells. This feature of tumor cells is commonly associated with acquired resistance to various kinds of treatments such as radio- or chemotherapy \cite{AneuCancer0, CancerCIN2, AneuCancer2,MS29}. Nonetheless, it is not clear whether aneuploidy contributes to and even drives tumor development or it is deleterious, since individuals carrying an extra copy of chromosome 21 have a 50\% lower probability of developing solid tumors than do individuals with the correct chromosome number \cite{MS18, MS19}. However, alterations in the karyotype, i.e. the canonical number and structure of chromosomes, represent the major cause of mental retardation and miscarriages \cite{MS1,MS2}. The presence of constitutional aneuploidy, with a proportion of aneuploid cell \textgreater 25\% in most tissues, is characteristic of Mosaic Variegated Aneuploidy syndrome (MVA), a rare autosomal disorder \cite{MS8}. Notably, MVA patients are affected by growth retardation, microcephaly and, among other developmental abnormalities, cancer predisposition \cite{MS9}.

Although aneuploidy is compatible with organism and cell viability, the presence of additional copies of chromosomes decreases the overall cellular fitness \cite {MS3}. However, gain or loss of specific chromosomes results in a small subset of cells with increased rate of transformation \cite {MS3,MS4}, with recent evidence showing that aneuploidy is likely to promote cancer development \cite{MS5, MS6}. In this scenario, aneuploidy could be a force driving cellular selection through a Darwinian process in which only cells with aneuploidy-driven adaptive traits, i.e. favourable changes in the dosage of specific sets of genes, overcome the microenvironmental challenges and survive \cite {MS7, Cahill}.

Since aneuploidy is generally associated with cancer, its potentially advantageous role in a physiological context has not been taken into consideration. In developmental biology, aneuploidy is regarded as an inescapable consequence of the rapid alternation fo S and M phases during embryo cleavage \cite {MS28} and has been associated to miscarriages and congenital defects. Counterintuitively, the percentage of cells undergoing DNA replication in solid tumors, which are mostly aneuploid, varies between 2 to 8\%, whereas a normal renewing epithelium such as the intestine exhibits a DNA replication index of approximately 16\% \cite {MS20}. Moreover, the majority of trisomies and chromosomal aberrations which are associated to miscarriages and birth defects, are known to originate during oogenesis, not during the cleavage stage, and are frequently due to non-disjunction or anaphase lagging in maternal meiosis I, that takes place mostly in the fetal life of the mother \cite {MS30}.
And yet, aneuploidies do exist that can be compatible with viable pregnancies and include those associated with chromosomes 13, 18, 21, X and Y \cite {MS31}. Defined threshold levels of aneuploidy that are compatible with life and yield mosaicism, are common in several mammals' embryos (\cite{MixoploAnimals} chapter 10), including humans \cite{ChaosEmbryo, HumanSC1}.

If our hypothesis of the existence of a diversification factor in stem cells is valid, a constitutional aneuploidy in the embryo would translate into widespread copy number variations in all human tissues \cite{AdultAndEmbryoMosaic}. Indeed, aneuploidy is well documented in the healthy adult human liver and brain, but is also detectable in skin, sperm and ovarian tissues from presumably normal individuals \cite{liver2, brain, AneuplTissues, AdultTissueCNV, MS56}. Two novel genome-wide association studies confirm the presence of unexpectedly high-frequency detectable levels of mosaicism in control human blood samples \cite{MS10,MS11} and even large CNVs have been recently detected 'within' tissues of the same individuals \cite{AdultTissueCNV}.

Genetic differences in monozygotic (MZ) twins (those who stem from the same zygote) represent an attractive model for studying somatic variations that occur during early embryonic development. Monozygotic twins frequently carry different copy number profiles \cite{MS13} and epigenetic marks \cite{MS14}. Using single nucleotide polymorphism arrays and fluorescent in situ hybridization analysis on a pair of monozygotic twins it has been found that one twin had monosomy X (45,X) in 7\% of proband nucleated blood cells, whereas the other twin had 45,X and 46,XY lineages, both present in 1\% of her cells \cite{MS15}.
In a separate study, a high incidence of segmental uniparental disomies, complete trisomies and several large copy number variants in multiple subjects was demonstrated. In one individual, five out of six alterations tested were detected in both blood and bladder tissue \cite{MS16}, indicating an early developmental origin. Taken together, these findings suggest that the resulting mosaicism in the adult partially originate in the embryonic stage, but a significant fraction also derives from {\it de novo} somatic modifications.

Occasionally the genome can be surprisingly tolerant to accommodate large copy number changes in apparently healthy subjects, raising the question whether this phenomenon only underlies a cell-replication error burden or it serves a physiological purpose and is instrumental to an increased plasticity and adaptability, which is fundamental to survive the continuous stream of environmental stresses.

Indeed a clear association exists between maternal age and miscarriage rates \cite {MS30} or between somatic mosaicism and ageing or cancer incidence \cite{MS10,MS11}. The positive correlation between age and elevated CNV profiles or clonal mosaicism holds true in MZ twins as well \cite{MS12, MS13, MS14}. The canonical interpretation of these scientific data is that, during their lifespan, cells and tissues undergo a series of mutational events and accumulate genetic abnormalities, ultimately leading to ageing and cancer. If this interpretation is correct, the presence of high CNV in most somatic tissues should translate into a default high probability of cancer development, especially so for highly proliferating tissues, but obviously this is not the case. On the contrary, if analyzed in the context of adaptability, the presence of an increasing somatic mosaicism and within-tissue CNV during lifespan, rather than a side-effect, is suggestive of a genetic diversification induced by environmental challenges during an individual's life \cite{MS65}. In fact, while germline genetic alterations are generally viewed as negative, a widespread somatic variation could be beneficial \cite {AneuploidyExtra}. For example, in tissues that frequently encounter pathogens, CNVs that eliminate viral receptors might enhance host survival.  Specific gains and losses of chromosomes harboring injury-resistance alleles in normal, nontransformed hepatocytes may render them differentially resistant to chronic insults such as viral hepatitis as well as alcohol- and fat-induced hepatitis \cite{liver}. Interestingly, mouse adult hepatocytes can increase and reduce their ploidy upon injury, thus resulting in a tremendous genetic heterogeneity and eventually leading to favorable cellular selection \cite{hepatocytesCIN}.

Studies in yeast showed that aneuploidy can provide a strong selective advantage in response to multiple environmental stressors \cite{Yeast, MS17}. Since chromosomes of higher eukaryotic genomes contain up to 12\% of genes arranged in functional neighborhoods, with a high level of gene co-expression \cite{MS21} and given that 5-10\%  of all genes are thought to be monoallelically expressed \cite{MS34}, even small changes in the total DNA content of a cell, i.e. low-level aneuploidy, are likely to cause phenotypic consequences \cite{MS65}. As an example, human developing brain has been proven to be a mosaic, with 30\% of the cells being aneuploid and up to 1.45\% frequency of aneuploidy per chromosome \cite {brain, MS32}. The resulting 10\% of healthy adult brain cells bearing abnormal chromosomal content could derive from the adaptive selection of three times the amount of aneuploid neuronal cells in the fetus and might explain the significant fraction of IQ-discordant monozygotic twins \cite {MS33}. Similarly, most of the  {\it de novo} CNVs in a set of provisional schizophrenia genes analyzed in MZ twins, have been shown to arise during developmental mitosis and are likely to account for the discordance in MZ twins for a variety of diseases including schizophrenia \cite{MS22}.

Embryonic stem cells are believed to be the primary source of somatic mosaicism \cite {AdultAndEmbryoMosaic}, indeed, although embryonic stem cells (ESCs) or induced pluripotent stem cells (iPSCs) accumulate specific chromosomal changes when cultured {\it in vitro} \cite{MS23}, these latter changes alone cannot account for the genetic heterogeneity displayed by these cells \cite {iPSC}. It has been recently published that as much as 30\% of the original normal fibroblast population from which iPSC were derived, presents somatic CNVs \cite{AdultMosaic1}, pointing to an ESC-specific genetic diversification program which is activated during development.
Despite this strong, but indirect evidence for a wide-spread mosaicism in the human stromal population of skin, human multipotent skin-derived precursor cells are known to keep stable karyotype in culture \cite {MS35}. Likewise, hematopoietic and gastric stem-like cells \cite {MS36, MS37} and Adipose-derived Stromal Cells \cite{AdultStemAneu} are known to maintain genetic stability in long-term culture. Recently, human cardiac stem cells have been isolated and they displayed long-term karyotipic stability in culture \cite{MS38}. In another study {\it in vitro} human-derived adult mesenchymal stromal cells have been shown to acquire detectable aneuploidy that is not related to culture and might be donor-dependent \cite{MS24}.

A recent comprehensive analysis of chromosomal abnormalities in cultured human adult stem cells revealed that the genetic identity of mesenchymal, neural and hematopoietic stem cell changes during cell culture passages and specific aberrations confer growth advantage in a cell lineage-specific manner \cite{MS25}. Conversely, previous studies confirmed the diploid nature of neural stem cells over 100 passages in culture \cite{MS39}.

Even though there is conflicting literature in this field, many cultured adult stem-like cells are reported to display genetic stability and the opposing results obtained by different groups might depend on the experimental settings, on the individuals from which adult stem cells were derived as well on the subset of adult stem cells.
Taken together, these observation point to a basal genetic stability of adult stem cell that is not compatible with the presence of extensive basal levels of CNV in adult dividing tissues such as liver, small intestine and pancreas \cite{AdultTissueCNV}. Remarkably, almost 80\% of the CNVs detected in \cite {AdultTissueCNV} are found in gene sequences, pointing to a non-random distribution of the genetic modifications.
If we assume that only ESCs have the potential to generate somatic mosaicism and that all somatic changes are age-related, then a consistent fraction of adult stem cells in renewing human organs should display the genetic marks of the transmitted CNV. Since most adult stem cells display normal karyotype, we envision a potential contribution of adult stem cells in physiologically and/or {\it adaptively} generating the somatic genetic heterogeneity that distinguishes adult tissues. In this regard, although ESCs and adult stem cells display strikingly different aneuploidy rates, they might both retain the ability to activate a molecular mechanism of genetic diversification in the presence of environmental stress. Therefore a chronic insult would result in abnormally sustained diversification, tremendously raising the probability that one or more aneuploid clones with tumorigenic properties would be positively selected.

As a matter of fact, it has been demonstrated that only a minor population of cells derived from a single cancer cell clone (diversified population from a single cell), could be responsible for drug resistance and, upon removal of the drug, the population spontaneously reverted to a sensitive state \cite {ReverseResist, MS40}. These evidences support the notion that tumor cells are endowed with a diversification potential. Indeed, clonal heterogeneity within tumors is the main cause of tumor dormancy and resistance to anti-cancer therapies \cite{MS29, Landau:2013}.

Many types of cancers partially inherit the hierarchical structure of the tissues they have derived from and are assumed to be propagated by stem-like tumor cells \cite{CancerStem1}. Therefore it is conceivable that cancer stem cells (CSCs) or any tumor-initiating cell employ increased aneuploidy rates actively, to the population advantage, to increase the overall fitness and adaptability of the tumor.

It is worth noting that, to date and to our knowledge, only two studies assessed the relationship between aneuploidy and CSCs \cite{MS26, MS27}. In the first paper, Kusumbe and Bapat evaluated the expression of stem-cell markers and the DNA content distribution of fluorescently labeled ovarian cancer cells after subcutaneous injection into immunodeficient mice. The authors found that, unlike label-free tumor cells, the label-retaining (quiescent) cells displayed stem-cell markers and were embedded with a small fraction of aneuploid cells. Treatment with chemotherapy increased the percentage of quiescent cells in the overall population and selectively stimulated the proliferation of the aneuploid fraction, which retained stemness properties upon removal of the drug \cite{MS26}. A second study by Fujimori et al., reveals again that stressful conditions favor the emergence of CSC-like clones from differentiating embryonic stem cells in {\it in vitro} culture \cite{MS27}.

We hypothesize that genetic heterogeneity is obtained in ESCs, adult stem cells and CSCs through increased aneuploidy rates and, by revisiting a recent work in Section 3, we suggest some evidence of the existence of a {\it diversification factor}.
In Section 4, in the context of a controlled diversification due to variable aneuploidy rates, we show how to interpret quasispecies theory to predict a maximal aneuploidy rate, an error threshold, after which a dominant cancer subpopulation looses its identity and therefore its fitness, what is referred to as error catastrophe \cite{quasispecies1}.

\section*{Evidence for Adaptive Aneuploidy Rate}

As already reviewed in Section 2, the development of single-cell sequencing techniques \cite{MS54} has recently allowed a wide range of studies analyzing variability in primary and metastatic tumor cells, as well as healthy tissues \cite{MS55,MS56}. In this section we would like to revisit and comment some specific experimental evidence in light of our proposal, focusing on several measures of aneuploidy rate, and showing how the assumption of a diversification factor can give new meaning to the variability observed in tumor sub-populations subject to distinct micro-environments.

The analysis in this section is phenomenological, meaning that we explain  and reinterpret existing experimental work, showing how the inner logic of our argumentations can severely constraint the causes and interpretation of the data we review. We chose to perform an in-depth analysis of a single recent study \cite{SingleCell}, so that the flow of our discourse is unified and made coherent by constant reference to the same context. At the same time, we support our arguments with related experimental works, when appropriate.

The main focus and objective of \cite{SingleCell} was to show that metastatic tumors are likely to be the product of single clones proliferation from the primary tumor by observing the microvariation of the integer copy number of consensus sequences in individual tumor cells through single-cell sequencing. A coarse ploidy distribution of a large number of cells from a breast tumor and one of its metastasis was plotted in (\cite{SingleCell} figure 3a-b) as an histogram with respect to the total DNA content. These ploidy distributions showed, for both primary and metastatic tumors, two peaks, one around twice and another at four times the total amount of DNA, this double peaked distribution accounted by the presence, in each tumor tissue sample, of roughly $50\%$ of normal diploid cells. Importantly, whereas in the primary tumor a significant fraction of the gated normal cells was pseudodiploid, in the paired metastasis the normal population was likely to derive from the stromal content of the tissue (\cite{SingleCell} figure 4).

The authors of \cite{SingleCell} performed very refined measurements of copy number profiles, across all chromosomes, from a small subset of cells (hundreds) in sections of primary and metastatic tumors and generated a neighbor-joining tree of these profiles. This analysis showed that metastatic and primary aneuploid cells were closely related in the neighbor-joining tree derived from the clustering, and yet they produced clearly distinct sub-clusters. The conclusion of this single cell study was that the metastasis proliferated from a single cell derived from the aneuploid subpopulation of the primary tumor, since no pseudodiploid cancer cells were observed in the metastatic tumor.

Other recent studies performed in myeloproliferative disorders \cite{MS48} and melanoma \cite{MS57}, kidney \cite{MS49, MS50} and pancreas \cite{MS51} tumors or, again, in breast cancer \cite{MS52}, arrived at similar conclusions, pointing to a late, metastasis-specific diversification of primary tumor-derived cells (\cite{MS46}, figure S14), \cite{MS47}. In \cite{MS50}, the presence of aneuploid cells in most primary tumors examined, is well documented by ploidy analyses in supplementary figure 10 and metastases show a marked increase in allelic imbalance as compared to primary tumor regions. The authors from  \cite{MS50} conclude that tumor heterogeneity is probably driven by aneuploidy and that chromosomal aberrations contribute substantially to genetic intratumor heterogeneity. Notably, as a further evidence of the robustness of our reasoning, even in an evolutionary context where the primary tumor and the metastasis share most of the sequenced regions, there is a striking variation in copy number specifically in the metastatic counterpart (\cite{MS57}, figures S5 to S11).

This concept is exemplified in \cite{SingleCell}, but here, both the coarse ploidy distribution analysis and the refined, single cell copy number count for primary and metastatic tumors, drive us to a more powerful conclusion: the genetic variability of the aneuploid clone in the metastasis is greater than its corresponding variability in the primary tumor from which it came. Even if we took into account a parallel progression model as opposed to a punctuated or linear evolution \cite{MS45, MS53}, the final result would not change. Regardless of whether the metastasis developed in a later, much shorter time than the primary tumor or whether its origins date back to the first stages of primary tumor dissemination, the metastatic population in the paper from Navin et al. has diversified more than its parental population in the primary tumor, even though the aneuploid cells compartment in the primary tumor analyzed by Navin et al. does not represent a minority of the population, but it has itself expanded considerably at some point during the tumor evolution history.

To justify our claim, we note that the Euclidean distances in the neighbor-joining tree for the aneuploid cells from the metastatic tumor studied in \cite{SingleCell} showed much greater variability than the Euclidean distances for the corresponding aneuploid cells in the primary tumor\footnote{These distances were calculated with respect to a common root profile, mutual distances among individual profiles are likely to be ever greater}. Granted that the cited study dealt with very small sample populations, but a closer inspection and analysis of the tightness of the variance of Euclidean distances in the primary tumor subpopulation, as opposed to the variance of Euclidean distances of the metastatic tumor subpopulation, would almost certainly reveal a statistically significant discrimination of the two.

The hypothesis of a larger variability of the copy number profiles of the metastatic subpopulation is clearly supported by a closer inspection of the tails of the ploidy distributions of primary and metastatic tumor populations analyzed in \cite{SingleCell}. The right-hand sides of the tetraploid peaks for the metastatic tumor have distinctly thicker and longer tails than the corresponding tetraploid peaks in the primary tumor, suggesting greater variability of aneuploidy in the former. Other works point to a similar conclusion, for example, it has been shown recently that, although sharing most of the examined somatic single-nucleotide variants, in vitro cultured low-passage melanoma cells have higher copy number variation when compared to the parental tumor \cite{MS58}.

Ploidy distributions, in their simplicity, offer even more scope for interpretation and testing of the hypothesis that cancer cells have the ability to self-regulate their aneuploidy rate. Indeed, a single cell clone could be capable of generating a diverse metastasis either because of inherent chromosomal instability, or because its rate of aneuploidy is somehow increased under the stress conditions of a new tissue embedding.

Let's assume first that the single clone from the primary tumor has chromosomal instability. The dispersion of metastatic cells should not be in any way preferential to such cells (even if their successful embedding in a tissue may be), so there will be a, possibly small, sub-population of cells in the primary tumor with similar or higher aneuploidy rate than the cell generating the metastasis. This subpopulation of the primary tumor, by its greater aneuploidy rate, will be more adaptable and likely self-sustaining, and it should be observable as a long tail in the ploidy distribution of the primary tumor population, the tail will be much thinner for the primary tumor than the metastatic tumor, since high aneuploidy rate cells are only a sub-population of the primary tumor. However, {\it no such long and thin tail is observed experimentally, for the primary tumor, in \cite{SingleCell}}. It is still possible, if highly unlikely, that the metastatic cell is an extreme outlier, with no comparable cells left in the primary tumor, but then we would see a much more pronounced evolutionary difference between primary and metastatic aneuploid populations than what is observed. The only logical conclusion is that the single metastatic cell clone was not essentially different from its primary population before starting to proliferate in the new environment.

This argument leaves only one other option: that some cancer cells are capable of altering and adapting their aneuploidy rate under stress, or under changes in the environment.

\section*{Aneuploid Error Threshold}

The concept of quasispecies was first introduced in \cite{Eigen0, Eigen1}, and it is a powerful way to relate the structure of population dynamics to the error rate of single base replication in viruses or unicellular organisms \cite{quasispecies1, quasispecies2}. The most important consequence of the theory is that it is possible to determine theoretically a threshold on the error rate such that, if the error rate of replication of genomic sequences is pushed above the threshold, the subpopulation will not be able to retain its identity, under mild conditions on subpopulations interactions and fitness distribution \cite{NewQS1}. Here we sketch the argument that leads to the error threshold inequality and we refer to \cite{quasispecies1, NewQS1, NewQS2} for details.

Assuming there are $N$ subpopulation types within a population, we start by writing down the differential equation that describes the rate of change $dx_m$ of type $m$ in terms of the instantaneous size $x_j(t)$, $j=1,...,N$ of all types:
\begin{equation}
\frac{dx_m}{dt}=(W_{mm}-\bar E(t))x_m(t)+\sum_{k\neq m} W_{ik}x_k(t),
\end{equation}
$W_{mm}$ is the rate of effective excess production of of sub-population type $m$, and if we consider the genetic sequence associated to type $m$, we can write $W_{mm}=Q_{mm}A_m-D_m$, with $Q_{mm}$ the probability of precise reproduction of sequence $m$, $A_m$ the growth rate of type $m$, and $D_m$ its mortality rate. $\bar E(t)=\sum_k E_k x_k$ is the average, over all types, of the excess reproduction rate, with $E_k=A_i-D_i$ and, finally, $W_{mk}$ is the rate of production of type $m$ by erroneous reproduction of type $k$.

Assuming a steady state in which ${dx_m}/{dt}=0$ and neglecting in first approximation the contributions $\sum_{k\neq m} W_{mk}x_k(t)$, it is possible to derive a condition that constraint the probability of precise reproduction of sequence $m$:
\begin{equation}
\bar \sigma_{m}Q_{mm}>1,
\end{equation}
where  $\bar \sigma_{m}=A_m/(D_m+\bar E_{k\neq m})$, with $\bar E_{k\neq m}=\sum_{k\neq m} E_k x_k$, is the average superiority of a master sequence associated to a dominant subpopulation versus competitor sequences, essentially, $\bar \sigma_m$ is an index of relative fitness \cite{quasispecies1}. If a master sequence has length $v_m$, and we denote by $\bar q$ the average fidelity of single nucleotide reproduction, then $Q_{mm}=\bar q^{v_m}$ and the error threshold can be written as
\begin{equation}
\frac{\ln \bar \sigma_m}{v_m}\geq1-\bar q.
\end{equation}
Remarkably, Eq. 3 establishes a phase transition on the information content, if the error rate of single nucleotide reproduction goes above $\frac{\ln \bar \sigma_m}{v_m}$ the information contained in the master sequence will disintegrate, in the sense that the loss of information in the sequence due to reproduction errors will not be compensated by a sufficiently high fitness relative to other subpopulations and the subpopulation associated to the master sequence will implode \cite{quasispecies1, NewQS1}.

In complex organisms, the quasispecies model is potentially applicable only in specific scenarios, such as competition among embryo stem cells during development, adult stem cells and progenitor cells proliferation, and, crucially, cancer cells, where subpopulations compete with each other under limited resources and changing environment. However, the applicability of the basic quasispecies model, originally devised in the setting of virus RNA replication, has been put into question as appropriate for eucaryotes and specifically for cancer cells. Eucaryotic cells reproduce semi-conservatively meaning that the parental double strand degenerates in the process of generating two daughter double strands, making the authors of \cite{QScancer1} raise the possibility that, for high enough replication error rates, the master sequence, seen here as the double strand of DNA, would eventually disappear, and they suggested more refined quasispecies models that take into consideration this phenomenon.

Even more seriously, the applicability of quasispecies theory to human cells is put into question by the exceedingly high size of the human genome as compared to RNA viruses. In fact, in order for the quasispecies not to undergo genetic drift, the neutral space around a fitness peak should be sufficiently small to be completely explored by the population. The complexity and the inherent mutational and phenotypical robustness of human genome amplifies its neutral space, preventing quasispecies evolution even at higher than normal mutation rates as in cancer cells \cite{NuetralSpace}. This fact, together with very low single nucleotide errors for humans, implies that the fitness of mutants of an hypothetical master sequence does not change significantly, and the fitness distribution of mutants around the master sequence is likely to decay linearly, or sub-linearly, a scenario under which no error threshold is possible \cite{NewQS1}.

Indeed, the existence of the Lynch syndrome or hereditary non-polyposis colorectal carcinomas (HNPCC), which are characterized by a higher risk of incidence of colon cancer, show the inability of the basic quasispecies theory to predict the maximum single nucleotide error that is viable for a tumor. HNPCC tumors, as well as all micrositellite instable (MSI) colon cancers, arise because of a break down of the mismatch repair mechanism \cite{MIN1, MIN2}. Therefore, MSI cancer cells display increased error rates of single nucleotide replication by 1 to 3 orders of magnitude, with respect to the baseline probability $1-\bar q$ to have a single nucleotide error in healthy cells, estimated for human genome to range between $10^{-9}$ and $10^{-10}$ (\cite{Alberts} page 271, \cite{DNARepl1, DNARepl2, MS41}).

Now recall that the human genome has roughly $3.2*10^9$ nucleotides (\cite{Alberts} page 206), and note that for organisms with very large genomes, the relative superiority $\bar \sigma_m$ of a master sequence associated to a given subpopulation cannot be very large (vis a vis other subpopulations), as any given mutation will only affect marginally its fitness \cite{quasispecies1}, and therefore $\bar \sigma_m\approx 1$. Given these numerical estimates, according to the error threshold inequality in Eq. 3, MSI tumors would fail to satisfy the error threshold inequality to such an extent that they should not even exist. This is true even if we restrict our attention, in defining the master sequence, to conserved DNA, i.e. the $5\%$ of the human genome that is known to be coding  and essential to cell function (see again \cite{Alberts}, page 206).

We believe that these inconsistences of the basic quasispecies model, when applied to human cells, completely disappear if we replace single nucleotide errors with aneuploidy errors. Notably, in all scenarios where quasispecies theory could potentially apply, i.e. stem and progenitor cells proliferation and cancer cells, aneuploidy rates far exceed single nucleotide error rates in frequency and impact on the cell, so that, effectively, the leading cause in the evolution of a population will be the aneuploidy error, rather than the single nucleotide error, that can be neglected, especially when the mismatch repair genes are intact as it happens in the overwhelming majority of cancers and all healthy stem cells.

We need now to reinterpret the notion of fidelity of reproduction of a sequence adequately to properly define error thresholds in the presence of aneuploidy. Since we can neglect nucleotide errors, we assume a faithful reproduction of the genetic material when the copy number of each chromosome in a sequence (two, for example, in a diploid cell) is kept constant during replication, both numerically (number of physically distinct whole chromosomes or fragments) and structurally (translocations, deletions and amplifications of DNA). Although complex aneuploidy landscapes may arise, characterized by concurrent numerical and structural chromosomal changes, most of the somatic copy-number alterations (SCNAs) frequently found in tumor cells involve whole chromosomes or whole-arms (25\% of the genome), with only 10\% of the cancer cell genome being affected by focal SCNAs \cite{AneuCancer3, AneuploidyULTRA1, AneuploidyULTRA2}.

Therefore, we define now {\it the chromosomal master sequence length $c_m$} of a cell as the total number of all whole and fragmented chromosomes in its nucleus, and we define the {\it aneuploid fidelity $\bar A_m$} as the average probability that each whole or fragmented chromosome is reproduced exactly once in cell division, with no gain or loss of sub-chromosomal regions.
In this aneuploid scenario, the chromosomal master sequence length $c_m$ can fluctuate depending on the number of aneuploid copies of whole chromosomes or fragments, and the underlying nucleotide sequence will clearly differ according to which chromosomes or individual genes are affected by copy-number alterations in each cell.
Although tumors vary widely in the number and type of copy number changes, most of these comprise low-level alterations and only a few genes reach more than $20$ copy numbers, mainly due to their oncogenic or drug-resistance functions \cite{SingleCell, MS42, MS43, MS44}.

Aneuploid events can cause large phenotypical variations \cite{MS65, AneuploidyExtra, AneuploidyLeap, Yeast}, even a single error leading to chromosomal loss or addition can have large effects, therefore the fitness distribution around a master sequence is expected to display a sharp decay from the master sequence peak, in line with the types of fitness distribution known to express the error threshold \cite{NewQS1}. At least for cancer cells, sub-populations are sharply defined in terms of their aneuploid profile, as evinced from the single cell analysis works \cite{SingleCell} commented in Section 3, this is further evidence of the strong concentration of fitness distributions  around a few chromosomal sequence types.

Mutants of the master sequence, generated by even a single aneuploid error, and individual cells belonging to other sub-populations, are exceedingly unlikely to be able to mutate into cells expressing the master sequence, since any additional (erroneous) chromosome copy is subject to a wide variety of further partial deletions/additions, and only very few of them would correspond to a return to the master sequence configuration. Essentially, we can assume that the contribution of cells belonging to other subpopulation types to the dynamical evolution of the master sequence subpopulation is very small, and this is exactly the condition that led to the error threshold in the first place, since Eq. 3 is derived as a limiting stationary behavior of an interacting family of subpopulations described by Eq. 1, where the rate of growth of each of them is weakly affected by the cross-mutations derived from the other subpopulations \cite{NewQS1,quasispecies1}.

Given these caveats regarding distribution of fitness for chromosomal master sequences in the presence of aneuploidy and regarding sub-population interactions, we reach the conclusion that quasispecies theory is indeed applicable to cancer and stem cells, but only in the context of aneuploid chromosomal master sequences, neglecting the underlying nucleotide errors.

We can now replace variables in the error threshold inequality Eq. 3 to take into account not only the variable length $c_m$ of the chromosomal master sequence associated to all whole and fragmented chromosomes, but also the correction to the probability $Q_{mm}$ of precise reproduction of a sequence that aneuploidy entails. The probability of precise reproduction of a specific sequence of chromosomes $m$ can be expressed as $Q_{mm}=\bar A_m^{c_m}$, and the aneuploid error threshold inequality can be written as $\bar \sigma_{m}\bar A_m^{c_m}>1$, which gives us an aneuploid error threshold inequality formally identical to Eq. 3:
\begin{equation}
\frac{\ln \bar \sigma_m}{c_m}\geq 1-\bar A_m,
\end{equation}
however, each term in this equation has drastically different orders of magnitudes than the threshold inequality for DNA or RNA master sequences. To start with, as already stressed above, the relative superiority $\bar \sigma_m$ will have considerable fluctuations, since the chromosomal master sequence is much shorter than a nucleotide sequence, and even small variations in copy numbers can effect large phenotypical variations.

At the same time, the diversity of subpopulations in primary tumors \cite{SingleCell, MS49, MS50, MS51, MS52, MS53} implies that, in a fully developed tumor, different subpopulations do not have extremely different relative superiority $\bar \sigma_m$, a scenario that would lead to a single, highly dominant subpopulation. It is therefore reasonable to assume at the very most $\bar \sigma_m\in [10^2,10^3]$, for the subpopulations of highest relative superiority. We know moreover that highly aneuploid tumors have higher fitness \cite{AneuploidyFit}, so larger values of $\bar \sigma_m$ are likely to be associated with large values of $c_m$,  up to the order of $10^2$ \cite{MS69, AneuploidyULTRA2, MS54, MS57}.

Write $E_m=1-\bar A_m$ in Eq. 4, with $E_m$ the aneuploidy error rate, i.e. the average probability that there is at least one new aneuploid defect for each chromosome or fragment of chromosome during cell replication. If we call $T(E_m)$ the threshold aneuploidy error rate above which a chromosomal master sequence is not viable, and if we take $\bar\sigma_m\in [10^2,10^3]$, $c_m\approx 10^2$, then Eq. 4 gives $T(E_m)\approx 10^{-2}$, which is consistent with the estimates of $E_m$ for cancer cells, in the range $[10^{-3}, 10^{-1}]$, derived in \cite{CancerCIN0, CancerCIN1, CancerCIN2, CancerCIN3, CancerCIN4}. Our argument implies that the more a cell is aneuploid, the tighter the error threshold bound is, and that highly aneuploid cancer cells, known to be most adaptable \cite{AneuploidyFit, MS29, CancerAneuploidyExtra, AneuCancer7}, are already working with aneuploid error rates close to the limit of a viable quasispecies.

There is some evidence that indeed aneuploidy rates in the tumor can affect the prognosis of cancer patients, in \cite{AneuCancer3} and \cite{AneuCancer6} it is suggested that a moderate tumor aneuploidy rate worsen the prognosis, while a very high aneuploidy rate is associated in \cite{AneuCancer3} with improved patient outcome, consistently with the quasispecies and error threshold catastrophe approach.

\section*{A Diversification Factor}

In Section 3 we inferred from some experimental results of \cite{SingleCell} and other published supportive information \cite{MS46, MS50, MS57}, that metastatic cancer cells have higher aneuploidy rates than the corresponding original subpopulation of the primary tumor and we concluded that this differential could only be explained by assuming an adaptive cellular response sensitive to changes in the environment.

Moreover, as pointed out in Section 1, high aneuploidy rates are common in cancer cells and embryo stem cells, and, to a much lower degree in some types of adult stem cells, showing again a fine-tuned
differential in the degree of diversification generated by aneuploidy rates. Taken together, these observations imply that adaptive aneuploidy is regulated through a specific cellular mechanism, a diversification factor, already active in embryo and adult stem cells to variable degrees, and then excessively reactivated in cancer cells.

The duration and degree of activation of a diversification factor, dependent on the phenotype of cultured cells and/or culture conditions and manipulation, might underlie the conflicting results in the measurement of aneuploidy in adult stem cell cultures (Section 1).
Since failed cytokines or multipolar mitosis (resulting in an increase or a reduction of chromosomes, respectively) cause numerical chromosome abnormalities, but also uniparental chromosome sets, it could be that controlled expression of protein(s) involved in determining/controlling the segregation of chromosomes, would result in a certain degree of aneuploidy in a population of cells. Bialletic mutation of the {\itshape BUB1B} gene results in constitutional aneuploidy  \cite{MS8} and a near-diploid aneuploidization has been experimentally induced by overexpression of the BUB1 protein, both genes coding for key proteins involved in the anaphase checkpoint machinery \cite{MS59}. Similarly, transient induction of the mitotic checkpoint gene {\itshape Mad2} leads to chromosomal instability and to tumor relapse \cite {MS66}.
Another component of the mammalian mitotic checkpoint, the Centromere-associated protein-E (CENP-E), is essential to prevent aneuploidy \cite{MS67}. Finally, in a recent paper, three new genes located on chromosome 18q, have been identified that contribute to suppress chromosomal instability and are subject to frequent copy number loss in colorectal cancers characterized by chromosomal instability \cite{MS68}. Therefore adaptive aneuploidy could be driven by timed and controlled gene expression of a specific protein or a set of proteins regulating the mitotic spindle assembly, centrosome duplication and cell-cycle checkpoints. As a matter of fact, inactivating mutations in the {\itshape STAG2} gene (encoding a component of the cohesin complex required for normal chromosome segregation) is known to induce aneuploidy \cite{MS5}.

Interestingly, DNA hypomethylation has been associated with aneuploidy and cancer and post-transcriptional silencing of the DNA-methyltransferases protein DNMT1 has recently proven to induce aneuploidy in fibroblasts and colon cancer cells \cite{MS60}. A Recent study focused on the high frequency mutations occuring at non-methylated cytosines throughout the genome of many breast cancers and found that overexpression of DNA cytosine deaminase APOBEC3B correlates with mutation rate \cite{Burns:2013}.

The mutator phenotype is a common feature of cancer cells \cite{Loeb:2008, Loeb:2011} and, although promoting the insurgence of driver mutations, many deleterious passenger mutations accumulate during tumor evolution \cite{McFarland:2013}. Buffering mechanisms such as the proteasome or the chaperone systems reduce the detrimental effects of passenger mutations and copy number alterations \cite{Burgess:2013}. In this regard it is worth citing the potentially relevant role of HSP90 in adaptive aneuploidy. HSP90 is a molecular chaperone conserved from plants to mammals, that assists metastable client proteins and helps them maintaining their correct conformation or refolding after mutational events or stress-induced denaturation \cite{MS62}. Due to its pleiotropic and fundamental function, HSP90 has been regarded as an evolutionary "capacitor", i.e. it buffers genetic variation (in terms of mutations or copy number) resulting in phenotypic robustness \cite {MS61}. Many developmental and metabolic phenotypes are threshold traits where the aberrant phenotype is triggered when the disease-associated factor falls below a crucial level, and this level may differ between species. The cryptic variants that under normal conditions are buffered by HSP90, would eventually be exposed and tested by natural selection in a stressful environment where HSP90 function is compromised \cite {MS63}. In such cases, HSP90 would act on the pre-stored genetic diversity, but it has recently been correlated to the insurgence of aneuploidy and consequently increased adaptability under stress conditions in a yeast model \cite{HeatShock}.

Due to the arrangement of eukaryotic genes in functional neighborhoods inside chromosomes and due to the high level of gene co-expression and monoallelic expression, even small changes in the total DNA content of a cell, i.e. low-level aneuploidy, are likely to result in a big phenotypic leap and enable the cell to explore a wide region of phenotypic landscapes. Thus, by altering the dosage of regulatory factors, aneuploidy can cause broad gene expression changes well beyond a direct DNA dosage effect. In relatively small populations under a strong selective force, the number of mutations with sufficient phenotypic effect to achieve adaptation is limited and increase in aneuploidy rate would certainly result in favorable cellular selection. Certain somatic evolutionary processes, such as the clonal expansion in early tumor/metastasis progression or relapse after drug treatment, may fall into this category. Therefore, the great ability of tumor metastases to resist therapeutic treatments is intimately connected to their higher copy number alterations as compared to primary tumors and this confirms our interpretation of the most recent literature.

Finally, starvation and related intercellular signaling, could be a possible direct or indirect trigger of adaptive aneuploidy because it is potentially at work in all organisms where variable aneuploidy is known to play a positive, adaptive effect, including plants, unicellular organisms and animals. Indeed, both replication stress and nucleotide deficiency is associated with genomic instability \cite{MS64, MS68}.
Elevated cell replication rates in the presence of less than optimal developmental conditions for the embryo would justify an increased microvariation of ploidy \cite{AdultAndEmbryoMosaic} to foster adaptation, and even in the adult organism several organs may be amenable to continuing adaptation through variable aneuploidy rates \cite{liver,brain, hepatocytesCIN}.
Our point is that only by taking a broad view that puts aneuploidy rate in its adaptive and evolutionary context, understanding its role across species and across states of embryo development, we can hope to identify the potential causes of adaptive aneuploidy in cancer.

A validation of the conjecture that cancer cells can adapt their aneuploidy rate through a diversification factor would have significant therapeutical implications, as it would provide a biological way, mostly inactive or less sensitive in healthy adult cells, to apply quasispecies error catastrophe strategies, along the lines of Section 4, to weaken cancer populations, a long held hope that may yet prove itself true.






\section*{Acknowledgments}
This paper is dedicated to Rosanna Gonzales, she never ceased to hope

\bibliography{CancerAneuploidyQuasiSpecies_References}

\begin{thebibliography}{100}
\providecommand{\url}[1]{\texttt{#1}}
\providecommand{\urlprefix}{URL }
\expandafter\ifx\csname urlstyle\endcsname\relax
  \providecommand{\doi}[1]{doi:\discretionary{}{}{}#1}\else
  \providecommand{\doi}{doi:\discretionary{}{}{}\begingroup
  \urlstyle{rm}\Url}\fi
\providecommand{\bibAnnoteFile}[1]{%
  \IfFileExists{#1}{\begin{quotation}\noindent\textsc{Key:} #1\\
  \textsc{Annotation:}\ \input{#1}\end{quotation}}{}}
\providecommand{\bibAnnote}[2]{%
  \begin{quotation}\noindent\textsc{Key:} #1\\
  \textsc{Annotation:}\ #2\end{quotation}}
\providecommand{\eprint}[2][]{\url{#2}}

\bibitem{AneuCancer0}
Torres EM, Williams BR, Amon A (2008) Aneuploidy: cells losing their balance.
\newblock Genetics 179: 737-46.
\bibAnnoteFile{AneuCancer0}

\bibitem{CancerCIN2}
Weaver BAA, Silk AD, Montagna C, Verdier-Pinard P, Cleveland DW (2007)
  Aneuploidy acts both oncogenically and as a tumor suppressor.
\newblock Cancer Cell 11: 25-36.
\bibAnnoteFile{CancerCIN2}

\bibitem{AneuCancer2}
Pavelka N, Rancati G, Li R (2010) Dr jekyll and mr hyde: role of aneuploidy in
  cellular adaptation and cancer.
\newblock Curr Opin Cell Biol 22: 809-15.
\bibAnnoteFile{AneuCancer2}

\bibitem{AneuCancer3}
McGranahan N, Burrell RA, Endesfelder D, Novelli MR, Swanton C (2012) Cancer
  chromosomal instability: therapeutic and diagnostic challenges.
\newblock EMBO Rep 13: 528-38.
\bibAnnoteFile{AneuCancer3}

\bibitem{PoliPloPlants}
te~Beest M, Le~Roux JJ, Richardson DM, Brysting AK, Suda J, et~al. (2012) The
  more the better? the role of polyploidy in facilitating plant invasions.
\newblock Ann Bot 109: 19-45.
\bibAnnoteFile{PoliPloPlants}

\bibitem{HeatShock}
Chen G, Bradford WD, Seidel CW, Li R (2012) Hsp90 stress potentiates rapid
  cellular adaptation through induction of aneuploidy.
\newblock Nature 482: 246-50.
\bibAnnoteFile{HeatShock}

\bibitem{MixoploAnimals}
Soom Av, Boerjan M (2002) Assessment of mammalian embryo quality: invasive and
  non-invasive techniques.
\newblock Dordrecht: Kluwer Academic Publishers.
\bibAnnoteFile{MixoploAnimals}

\bibitem{ChaosEmbryo}
Ledbetter DH (2009) Chaos in the embryo.
\newblock Nat Med 15: 490-1.
\bibAnnoteFile{ChaosEmbryo}

\bibitem{HumanSC1}
Peterson SE, Westra JW, Rehen SK, Young H, Bushman DM, et~al. (2011) Normal
  human pluripotent stem cell lines exhibit pervasive mosaic aneuploidy.
\newblock PLoS One 6: e23018.
\bibAnnoteFile{HumanSC1}

\bibitem{PigsEmbryos}
Hornak M, Oracova E, Hulinska P, Urbankova L, Rubes J (2012) Aneuploidy
  detection in pigs using comparative genomic hybridization: from the oocytes
  to blastocysts.
\newblock PLoS One 7: e30335.
\bibAnnoteFile{PigsEmbryos}

\bibitem{liver}
Duncan AW, Hanlon~Newell AE, Bi W, Finegold MJ, Olson SB, et~al. (2012)
  Aneuploidy as a mechanism for stress-induced liver adaptation.
\newblock J Clin Invest 122: 3307-15.
\bibAnnoteFile{liver}

\bibitem{brain}
Rehen SK, Yung YC, McCreight MP, Kaushal D, Yang AH, et~al. (2005)
  Constitutional aneuploidy in the normal human brain.
\newblock J Neurosci 25: 2176-80.
\bibAnnoteFile{brain}

\bibitem{AdultAndEmbryoMosaic}
Liang Q, Conte N, Skarnes WC, Bradley A (2008) Extensive genomic copy number
  variation in embryonic stem cells.
\newblock Proc Natl Acad Sci U S A 105: 17453-6.
\bibAnnoteFile{AdultAndEmbryoMosaic}

\bibitem{CancerStem1}
Nguyen LV, Vanner R, Dirks P, Eaves CJ (2012) Cancer stem cells: an evolving
  concept.
\newblock Nat Rev Cancer 12: 133-43.
\bibAnnoteFile{CancerStem1}

\bibitem{AdultMosaic1}
Abyzov A, Mariani J, Palejev D, Zhang Y, Haney MS, et~al. (2012) Somatic copy
  number mosaicism in human skin revealed by induced pluripotent stem cells.
\newblock Nature 492: 438-42.
\bibAnnoteFile{AdultMosaic1}

\bibitem{AdultStemAneu}
Grimes BR, Steiner CM, Merfeld-Clauss S, Traktuev DO, Smith D, et~al. (2009)
  Interphase fish demonstrates that human adipose stromal cells maintain a high
  level of genomic stability in long-term culture.
\newblock Stem Cells Dev 18: 717-24.
\bibAnnoteFile{AdultStemAneu}

\bibitem{hepatocytesCIN}
Duncan AW, Taylor MH, Hickey RD, Hanlon~Newell AE, Lenzi ML, et~al. (2010) The
  ploidy conveyor of mature hepatocytes as a source of genetic variation.
\newblock Nature 467: 707-10.
\bibAnnoteFile{hepatocytesCIN}

\bibitem{AdultTissueCNV}
O'Huallachain M, Karczewski KJ, Weissman SM, Urban AE, Snyder MP (2012)
  Extensive genetic variation in somatic human tissues.
\newblock Proc Natl Acad Sci U S A 109: 18018-23.
\bibAnnoteFile{AdultTissueCNV}

\bibitem{quasispeciesBiology}
Ojosnegros S, Perales C, Mas A, Domingo E (2011) Quasispecies as a matter of
  fact: viruses and beyond.
\newblock Virus Res 162: 203-15.
\bibAnnoteFile{quasispeciesBiology}

\bibitem{quasispecies1}
Biebricher CK, Eigen M (2005) The error threshold.
\newblock Virus Res 107: 117-27.
\bibAnnoteFile{quasispecies1}

\bibitem{QScancer1}
Brumer Y, Michor F, Shakhnovich EI (2006) Genetic instability and the
  quasispecies model.
\newblock J Theor Biol 241: 216-22.
\bibAnnoteFile{QScancer1}

\bibitem{QScancer2}
Sol{\'e} RV, Deisboeck TS (2004) An error catastrophe in cancer?
\newblock J Theor Biol 228: 47-54.
\bibAnnoteFile{QScancer2}

\bibitem{QScancer3}
Diego D, Calvo GF, P{\'e}rez-Garc{\'\i}a VM (2012) Modeling the connection
  between primary and metastatic tumors.
\newblock J Math Biol .
\bibAnnoteFile{QScancer3}

\bibitem{QScancer4}
Itan E, Tannenbaum E (2012) Effect of chromosomal instability on the
  mutation-selection balance in unicellular populations.
\newblock PLoS One 7: e26513.
\bibAnnoteFile{QScancer4}

\bibitem{QScancer5}
Iwami S, Haeno H, Michor F (2012) A race between tumor immunoescape and genome
  maintenance selects for optimum levels of (epi)genetic instability.
\newblock PLoS Comput Biol 8: e1002370.
\bibAnnoteFile{QScancer5}

\bibitem{QSNowak1}
Michor F, Iwasa Y, Nowak MA (2004) Dynamics of cancer progression.
\newblock Nat Rev Cancer 4: 197-205.
\bibAnnoteFile{QSNowak1}

\bibitem{QSNowak2}
Nowak MA, Michor F, Iwasa Y (2006) Genetic instability and clonal expansion.
\newblock J Theor Biol 241: 26-32.
\bibAnnoteFile{QSNowak2}

\bibitem{MS28}
Cooper GM (1997) The cell: a molecular approach.
\newblock Washington, D.C.: ASM Press.
\bibAnnoteFile{MS28}

\bibitem{MS29}
Kreso A, O'Brien CA, van Galen P, Gan OI, Notta F, et~al. (2013) Variable
  clonal repopulation dynamics influence chemotherapy response in colorectal
  cancer.
\newblock Science 339: 543-8.
\bibAnnoteFile{MS29}

\bibitem{MS18}
Hasle H, Clemmensen IH, Mikkelsen M (2000) Risks of leukaemia and solid tumours
  in individuals with down's syndrome.
\newblock Lancet 355: 165-9.
\bibAnnoteFile{MS18}

\bibitem{MS19}
Satg{\'e} D, Sasco AJ, Lacour B (2003) Are solid tumours different in children
  with down's syndrome?
\newblock Int J Cancer 106: 297-8.
\bibAnnoteFile{MS19}

\bibitem{MS1}
Pai GS, Lewandowski RC, Borgaonkar DS (2003) Handbook of chromosomal syndromes.
\newblock New York: J. Wiley.
\bibAnnoteFile{MS1}

\bibitem{MS2}
Gropp A, Kolbus U, Giers D (1975) Systematic approach to the study of trisomy
  in the mouse. ii.
\newblock Cytogenet Cell Genet 14: 42-62.
\bibAnnoteFile{MS2}

\bibitem{MS8}
Hanks S, Coleman K, Reid S, Plaja A, Firth H, et~al. (2004) Constitutional
  aneuploidy and cancer predisposition caused by biallelic mutations in bub1b.
\newblock Nat Genet 36: 1159-61.
\bibAnnoteFile{MS8}

\bibitem{MS9}
Snape K, Hanks S, Ruark E, Barros-N{\'u}{\~n}ez P, Elliott A, et~al. (2011)
  Mutations in cep57 cause mosaic variegated aneuploidy syndrome.
\newblock Nat Genet 43: 527-9.
\bibAnnoteFile{MS9}

\bibitem{MS3}
Williams BR, Prabhu VR, Hunter KE, Glazier CM, Whittaker CA, et~al. (2008)
  Aneuploidy affects proliferation and spontaneous immortalization in mammalian
  cells.
\newblock Science 322: 703-9.
\bibAnnoteFile{MS3}

\bibitem{MS4}
Hernando E (2008) Cancer. aneuploidy advantages?
\newblock Science 322: 692-3.
\bibAnnoteFile{MS4}

\bibitem{MS5}
Solomon DA, Kim T, Diaz-Martinez LA, Fair J, Elkahloun AG, et~al. (2011)
  Mutational inactivation of stag2 causes aneuploidy in human cancer.
\newblock Science 333: 1039-43.
\bibAnnoteFile{MS5}

\bibitem{MS6}
Sheltzer JM, Blank HM, Pfau SJ, Tange Y, George BM, et~al. (2011) Aneuploidy
  drives genomic instability in yeast.
\newblock Science 333: 1026-30.
\bibAnnoteFile{MS6}

\bibitem{MS7}
Kolodner RD, Cleveland DW, Putnam CD (2011) Cancer. aneuploidy drives a mutator
  phenotype in cancer.
\newblock Science 333: 942-3.
\bibAnnoteFile{MS7}

\bibitem{Cahill}
Cahill DP, Kinzler KW, Vogelstein B, Lengauer C (1999) Genetic instability and
  darwinian selection in tumours.
\newblock Trends Cell Biol 9: M57-60.
\bibAnnoteFile{Cahill}

\bibitem{MS20}
Tannock I, Hill RP (1987) The Basic science of oncology.
\newblock New York: Pergamon Press.
\bibAnnoteFile{MS20}

\bibitem{MS30}
Hassold T, Hall H, Hunt P (2007) The origin of human aneuploidy: where we have
  been, where we are going.
\newblock Hum Mol Genet 16 Spec No. 2: R203-8.
\bibAnnoteFile{MS30}

\bibitem{MS31}
Ambartsumyan G, Clark AT (2008) Aneuploidy and early human embryo development.
\newblock Hum Mol Genet 17: R10-5.
\bibAnnoteFile{MS31}

\bibitem{liver2}
Duncan AW, Hanlon~Newell AE, Smith L, Wilson EM, Olson SB, et~al. (2012)
  Frequent aneuploidy among normal human hepatocytes.
\newblock Gastroenterology 142: 25-8.
\bibAnnoteFile{liver2}

\bibitem{AneuplTissues}
Iourov IY, Vorsanova SG, Yurov YB (2008) Chromosomal mosaicism goes global.
\newblock Mol Cytogenet 1: 26.
\bibAnnoteFile{AneuplTissues}

\bibitem{MS56}
Wang J, Fan HC, Behr B, Quake SR (2012) Genome-wide single-cell analysis of
  recombination activity and de novo mutation rates in human sperm.
\newblock Cell 150: 402-12.
\bibAnnoteFile{MS56}

\bibitem{MS10}
Jacobs KB, Yeager M, Zhou W, Wacholder S, Wang Z, et~al. (2012) Detectable
  clonal mosaicism and its relationship to aging and cancer.
\newblock Nat Genet 44: 651-8.
\bibAnnoteFile{MS10}

\bibitem{MS11}
Laurie CC, Laurie CA, Rice K, Doheny KF, Zelnick LR, et~al. (2012) Detectable
  clonal mosaicism from birth to old age and its relationship to cancer.
\newblock Nat Genet 44: 642-50.
\bibAnnoteFile{MS11}

\bibitem{MS13}
Bruder CEG, Piotrowski A, Gijsbers AACJ, Andersson R, Erickson S, et~al. (2008)
  Phenotypically concordant and discordant monozygotic twins display different
  dna copy-number-variation profiles.
\newblock Am J Hum Genet 82: 763-71.
\bibAnnoteFile{MS13}

\bibitem{MS14}
Fraga MF, Ballestar E, Paz MF, Ropero S, Setien F, et~al. (2005) Epigenetic
  differences arise during the lifetime of monozygotic twins.
\newblock Proc Natl Acad Sci U S A 102: 10604-9.
\bibAnnoteFile{MS14}

\bibitem{MS15}
Razzaghian HR, Shahi MH, Forsberg LA, de~St{\aa}hl TD, Absher D, et~al. (2010)
  Somatic mosaicism for chromosome x and y aneuploidies in monozygotic twins
  heterozygous for sickle cell disease mutation.
\newblock Am J Med Genet A 152A: 2595-8.
\bibAnnoteFile{MS15}

\bibitem{MS16}
Rodr{\'\i}guez-Santiago B, Malats N, Rothman N, Armengol L, Garcia-Closas M,
  et~al. (2010) Mosaic uniparental disomies and aneuploidies as large
  structural variants of the human genome.
\newblock Am J Hum Genet 87: 129-38.
\bibAnnoteFile{MS16}

\bibitem{MS12}
Forsberg LA, Rasi C, Razzaghian HR, Pakalapati G, Waite L, et~al. (2012)
  Age-related somatic structural changes in the nuclear genome of human blood
  cells.
\newblock Am J Hum Genet 90: 217-28.
\bibAnnoteFile{MS12}

\bibitem{MS65}
Chen G, Rubinstein B, Li R (2012) Whole chromosome aneuploidy: big mutations
  drive adaptation by phenotypic leap.
\newblock Bioessays 34: 893-900.
\bibAnnoteFile{MS65}

\bibitem{AneuploidyExtra}
Sheltzer JM, Amon A (2011) The aneuploidy paradox: costs and benefits of an
  incorrect karyotype.
\newblock Trends Genet 27: 446-53.
\bibAnnoteFile{AneuploidyExtra}

\bibitem{Yeast}
Pavelka N, Rancati G, Zhu J, Bradford WD, Saraf A, et~al. (2010) Aneuploidy
  confers quantitative proteome changes and phenotypic variation in budding
  yeast.
\newblock Nature 468: 321-5.
\bibAnnoteFile{Yeast}

\bibitem{MS17}
Rancati G, Pavelka N, Fleharty B, Noll A, Trimble R, et~al. (2008) Aneuploidy
  underlies rapid adaptive evolution of yeast cells deprived of a conserved
  cytokinesis motor.
\newblock Cell 135: 879-93.
\bibAnnoteFile{MS17}

\bibitem{MS21}
Al-Shahrour F, Minguez P, Marqu{\'e}s-Bonet T, Gazave E, Navarro A, et~al.
  (2010) Selection upon genome architecture: conservation of functional
  neighborhoods with changing genes.
\newblock PLoS Comput Biol 6: e1000953.
\bibAnnoteFile{MS21}

\bibitem{MS34}
Chess A (2012) Mechanisms and consequences of widespread random monoallelic
  expression.
\newblock Nat Rev Genet 13: 421-8.
\bibAnnoteFile{MS34}

\bibitem{MS32}
Dierssen M, Herault Y, Estivill X (2009) Aneuploidy: from a physiological
  mechanism of variance to down syndrome.
\newblock Physiol Rev 89: 887-920.
\bibAnnoteFile{MS32}

\bibitem{MS33}
Ehli EA, Abdellaoui A, Hu Y, Hottenga JJ, Kattenberg M, et~al. (2012) De novo
  and inherited cnvs in mz twin pairs selected for discordance and concordance
  on attention problems.
\newblock Eur J Hum Genet 20: 1037-43.
\bibAnnoteFile{MS33}

\bibitem{MS22}
Maiti S, Kumar KHBG, Castellani CA, O'Reilly R, Singh SM (2011) Ontogenetic de
  novo copy number variations (cnvs) as a source of genetic individuality:
  studies on two families with mzd twins for schizophrenia.
\newblock PLoS One 6: e17125.
\bibAnnoteFile{MS22}

\bibitem{MS23}
Lund RJ, N{\"a}rv{\"a} E, Lahesmaa R (2012) Genetic and epigenetic stability of
  human pluripotent stem cells.
\newblock Nat Rev Genet 13: 732-44.
\bibAnnoteFile{MS23}

\bibitem{iPSC}
Young MA, Larson DE, Sun CW, George DR, Ding L, et~al. (2012) Background
  mutations in parental cells account for most of the genetic heterogeneity of
  induced pluripotent stem cells.
\newblock Cell Stem Cell 10: 570-82.
\bibAnnoteFile{iPSC}

\bibitem{MS35}
Toma JG, McKenzie IA, Bagli D, Miller FD (2005) Isolation and characterization
  of multipotent skin-derived precursors from human skin.
\newblock Stem Cells 23: 727-37.
\bibAnnoteFile{MS35}

\bibitem{MS36}
Katayama Y, Miyamoto K, Takenaka K, Imajyo K, Shinagawa K, et~al. (2001)
  Chromosome analysis after ex vivo expansion of cd34(+) cells from human cord
  blood.
\newblock Cancer Genet Cytogenet 125: 161-2.
\bibAnnoteFile{MS36}

\bibitem{MS37}
Yang YC, Wang SW, Hung HY, Chang CC, Wu IC, et~al. (2007) Isolation and
  characterization of human gastric cell lines with stem cell phenotypes.
\newblock J Gastroenterol Hepatol 22: 1460-8.
\bibAnnoteFile{MS37}

\bibitem{MS38}
He JQ, Vu DM, Hunt G, Chugh A, Bhatnagar A, et~al. (2011) Human cardiac stem
  cells isolated from atrial appendages stably express c-kit.
\newblock PLoS One 6: e27719.
\bibAnnoteFile{MS38}

\bibitem{MS24}
Tarte K, Gaillard J, Lataillade JJ, Fouillard L, Becker M, et~al. (2010)
  Clinical-grade production of human mesenchymal stromal cells: occurrence of
  aneuploidy without transformation.
\newblock Blood 115: 1549-53.
\bibAnnoteFile{MS24}

\bibitem{MS25}
Ben-David U, Mayshar Y, Benvenisty N (2011) Large-scale analysis reveals
  acquisition of lineage-specific chromosomal aberrations in human adult stem
  cells.
\newblock Cell Stem Cell 9: 97-102.
\bibAnnoteFile{MS25}

\bibitem{MS39}
Sun Y, Pollard S, Conti L, Toselli M, Biella G, et~al. (2008) Long-term
  tripotent differentiation capacity of human neural stem (ns) cells in
  adherent culture.
\newblock Mol Cell Neurosci 38: 245-58.
\bibAnnoteFile{MS39}

\bibitem{ReverseResist}
Sharma SV, Lee DY, Li B, Quinlan MP, Takahashi F, et~al. (2010) A
  chromatin-mediated reversible drug-tolerant state in cancer cell
  subpopulations.
\newblock Cell 141: 69-80.
\bibAnnoteFile{ReverseResist}

\bibitem{MS40}
Workman P, Travers J (2010) Cancer: drug-tolerant insurgents.
\newblock Nature 464: 844-5.
\bibAnnoteFile{MS40}

\bibitem{Landau:2013}
Landau DA, Carter SL, Stojanov P, McKenna A, Stevenson K, et~al. (2013)
  Evolution and impact of subclonal mutations in chronic lymphocytic leukemia.
\newblock Cell 152: 714-26.
\bibAnnoteFile{Landau:2013}

\bibitem{MS26}
Kusumbe AP, Bapat SA (2009) Cancer stem cells and aneuploid populations within
  developing tumors are the major determinants of tumor dormancy.
\newblock Cancer Res 69: 9245-53.
\bibAnnoteFile{MS26}

\bibitem{MS27}
Fujimori H, Shikanai M, Teraoka H, Masutani M, Yoshioka Ki (2012) Induction of
  cancerous stem cells during embryonic stem cell differentiation.
\newblock J Biol Chem 287: 36777-91.
\bibAnnoteFile{MS27}

\bibitem{MS54}
Zong C, Lu S, Chapman AR, Xie XS (2012) Genome-wide detection of
  single-nucleotide and copy-number variations of a single human cell.
\newblock Science 338: 1622-6.
\bibAnnoteFile{MS54}

\bibitem{MS55}
Owens B (2012) Genomics: The single life.
\newblock Nature 491: 27-9.
\bibAnnoteFile{MS55}

\bibitem{SingleCell}
Navin N, Kendall J, Troge J, Andrews P, Rodgers L, et~al. (2011) Tumour
  evolution inferred by single-cell sequencing.
\newblock Nature 472: 90-4.
\bibAnnoteFile{SingleCell}

\bibitem{MS48}
Hou Y, Song L, Zhu P, Zhang B, Tao Y, et~al. (2012) Single-cell exome
  sequencing and monoclonal evolution of a jak2-negative myeloproliferative
  neoplasm.
\newblock Cell 148: 873-85.
\bibAnnoteFile{MS48}

\bibitem{MS57}
Turajlic S, Furney SJ, Lambros MB, Mitsopoulos C, Kozarewa I, et~al. (2012)
  Whole genome sequencing of matched primary and metastatic acral melanomas.
\newblock Genome Res 22: 196-207.
\bibAnnoteFile{MS57}

\bibitem{MS49}
Xu X, Hou Y, Yin X, Bao L, Tang A, et~al. (2012) Single-cell exome sequencing
  reveals single-nucleotide mutation characteristics of a kidney tumor.
\newblock Cell 148: 886-95.
\bibAnnoteFile{MS49}

\bibitem{MS50}
Gerlinger M, Rowan AJ, Horswell S, Larkin J, Endesfelder D, et~al. (2012)
  Intratumor heterogeneity and branched evolution revealed by multiregion
  sequencing.
\newblock N Engl J Med 366: 883-92.
\bibAnnoteFile{MS50}

\bibitem{MS51}
Campbell PJ, Yachida S, Mudie LJ, Stephens PJ, Pleasance ED, et~al. (2010) The
  patterns and dynamics of genomic instability in metastatic pancreatic cancer.
\newblock Nature 467: 1109-13.
\bibAnnoteFile{MS51}

\bibitem{MS52}
Shah SP, Roth A, Goya R, Oloumi A, Ha G, et~al. (2012) The clonal and
  mutational evolution spectrum of primary triple-negative breast cancers.
\newblock Nature 486: 395-9.
\bibAnnoteFile{MS52}

\bibitem{MS46}
Wu X, Northcott PA, Dubuc A, Dupuy AJ, Shih DJH, et~al. (2012) Clonal selection
  drives genetic divergence of metastatic medulloblastoma.
\newblock Nature 482: 529-33.
\bibAnnoteFile{MS46}

\bibitem{MS47}
Clifford SC (2012) Cancer genetics: Evolution after tumour spread.
\newblock Nature 482: 481-2.
\bibAnnoteFile{MS47}

\bibitem{MS45}
Klein CA (2009) Parallel progression of primary tumours and metastases.
\newblock Nat Rev Cancer 9: 302-12.
\bibAnnoteFile{MS45}

\bibitem{MS53}
Caldas C (2012) Cancer sequencing unravels clonal evolution.
\newblock Nat Biotechnol 30: 408-10.
\bibAnnoteFile{MS53}

\bibitem{MS58}
Parker SCJ, Gartner J, Cardenas-Navia I, Wei X, Ozel~Abaan H, et~al. (2012)
  Mutational signatures of de-differentiation in functional non-coding regions
  of melanoma genomes.
\newblock PLoS Genet 8: e1002871.
\bibAnnoteFile{MS58}

\bibitem{Eigen0}
Eigen M (1971) Selforganization of matter and the evolution of biological
  macromolecules.
\newblock Naturwissenschaften 58: 465-523.
\bibAnnoteFile{Eigen0}

\bibitem{Eigen1}
Eigen M, Schuster P (1979) The hypercycle, a principle of natural
  self-organization.
\newblock Berlin: Springer-Verlag.
\bibAnnoteFile{Eigen1}

\bibitem{quasispecies2}
Biebricher CK, Eigen M (2006) What is a quasispecies?
\newblock Curr Top Microbiol Immunol 299: 1-31.
\bibAnnoteFile{quasispecies2}

\bibitem{NewQS1}
Schuster P (2011) Mathematical modeling of evolution. solved and open problems.
\newblock Theory Biosci 130: 71-89.
\bibAnnoteFile{NewQS1}

\bibitem{NewQS2}
Eigen M, McCaskill JS, Schuster P (1989) The molecular quasi-species.
\newblock Advances in Chemical Physics 75: 149-263.
\bibAnnoteFile{NewQS2}

\bibitem{NuetralSpace}
Jenkins GM, Worobey M, Woelk CH, Holmes EC (2001) Evidence for the
  non-quasispecies evolution of rna viruses [corrected].
\newblock Mol Biol Evol 18: 987-94.
\bibAnnoteFile{NuetralSpace}

\bibitem{MIN1}
Vilar E, Gruber SB (2010) Microsatellite instability in colorectal cancer-the
  stable evidence.
\newblock Nat Rev Clin Oncol 7: 153-62.
\bibAnnoteFile{MIN1}

\bibitem{MIN2}
de~la Chapelle A, Hampel H (2010) Clinical relevance of microsatellite
  instability in colorectal cancer.
\newblock J Clin Oncol 28: 3380-7.
\bibAnnoteFile{MIN2}

\bibitem{Alberts}
Alberts B, Wilson JH, Hunt T (2008) Molecular biology of the cell.
\newblock New York: Garland Science, 5th ed., reference ed edition.
\bibAnnoteFile{Alberts}

\bibitem{DNARepl1}
Gundry M, Vijg J (2012) Direct mutation analysis by high-throughput sequencing:
  from germline to low-abundant, somatic variants.
\newblock Mutat Res 729: 1-15.
\bibAnnoteFile{DNARepl1}

\bibitem{DNARepl2}
Lange SS, Takata Ki, Wood RD (2011) Dna polymerases and cancer.
\newblock Nat Rev Cancer 11: 96-110.
\bibAnnoteFile{DNARepl2}

\bibitem{MS41}
Jiricny J (2006) The multifaceted mismatch-repair system.
\newblock Nat Rev Mol Cell Biol 7: 335-46.
\bibAnnoteFile{MS41}

\bibitem{AneuploidyULTRA1}
Gordon DJ, Resio B, Pellman D (2012) Causes and consequences of aneuploidy in
  cancer.
\newblock Nat Rev Genet 13: 189-203.
\bibAnnoteFile{AneuploidyULTRA1}

\bibitem{AneuploidyULTRA2}
Beroukhim R, Mermel CH, Porter D, Wei G, Raychaudhuri S, et~al. (2010) The
  landscape of somatic copy-number alteration across human cancers.
\newblock Nature 463: 899-905.
\bibAnnoteFile{AneuploidyULTRA2}

\bibitem{MS42}
Albertson DG (2006) Gene amplification in cancer.
\newblock Trends Genet 22: 447-55.
\bibAnnoteFile{MS42}

\bibitem{MS43}
Gajduskova P, Snijders AM, Kwek S, Roydasgupta R, Fridlyand J, et~al. (2007)
  Genome position and gene amplification.
\newblock Genome Biol 8: R120.
\bibAnnoteFile{MS43}

\bibitem{MS44}
Santarius T, Shipley J, Brewer D, Stratton MR, Cooper CS (2010) A census of
  amplified and overexpressed human cancer genes.
\newblock Nat Rev Cancer 10: 59-64.
\bibAnnoteFile{MS44}

\bibitem{AneuploidyLeap}
Rasnick D, Duesberg PH (1999) How aneuploidy affects metabolic control and
  causes cancer.
\newblock Biochem J 340 ( Pt 3): 621-30.
\bibAnnoteFile{AneuploidyLeap}

\bibitem{AneuploidyFit}
Bakhoum SF, Compton DA (2012) Chromosomal instability and cancer: a complex
  relationship with therapeutic potential.
\newblock J Clin Invest 122: 1138-43.
\bibAnnoteFile{AneuploidyFit}

\bibitem{MS69}
Vogt N, Lef{\`e}vre SH, Apiou F, Dutrillaux AM, C{\"o}r A, et~al. (2004)
  Molecular structure of double-minute chromosomes bearing amplified copies of
  the epidermal growth factor receptor gene in gliomas.
\newblock Proc Natl Acad Sci U S A 101: 11368-73.
\bibAnnoteFile{MS69}

\bibitem{CancerCIN0}
Lengauer C, Kinzler KW, Vogelstein B (1997) Genetic instability in colorectal
  cancers.
\newblock Nature 386: 623-7.
\bibAnnoteFile{CancerCIN0}

\bibitem{CancerCIN1}
Rajagopalan H, Nowak MA, Vogelstein B, Lengauer C (2003) The significance of
  unstable chromosomes in colorectal cancer.
\newblock Nat Rev Cancer 3: 695-701.
\bibAnnoteFile{CancerCIN1}

\bibitem{CancerCIN3}
Burns EM, Christopoulou L, Corish P, Tyler-Smith C (1999) Quantitative
  measurement of mammalian chromosome mitotic loss rates using the green
  fluorescent protein.
\newblock J Cell Sci 112 ( Pt 16): 2705-14.
\bibAnnoteFile{CancerCIN3}

\bibitem{CancerCIN4}
Camps J, Ponsa I, Ribas M, Prat E, Egozcue J, et~al. (2005) Comprehensive
  measurement of chromosomal instability in cancer cells: combination of
  fluorescence in situ hybridization and cytokinesis-block micronucleus assay.
\newblock FASEB J 19: 828-30.
\bibAnnoteFile{CancerCIN4}

\bibitem{CancerAneuploidyExtra}
Lee AJX, Endesfelder D, Rowan AJ, Walther A, Birkbak NJ, et~al. (2011)
  Chromosomal instability confers intrinsic multidrug resistance.
\newblock Cancer Res 71: 1858-70.
\bibAnnoteFile{CancerAneuploidyExtra}

\bibitem{AneuCancer7}
Duesberg P, Stindl R, Hehlmann R (2001) Origin of multidrug resistance in cells
  with and without multidrug resistance genes: chromosome reassortments
  catalyzed by aneuploidy.
\newblock Proc Natl Acad Sci U S A 98: 11283-8.
\bibAnnoteFile{AneuCancer7}

\bibitem{AneuCancer6}
Bakhoum SF, Danilova OV, Kaur P, Levy NB, Compton DA (2011) Chromosomal
  instability substantiates poor prognosis in patients with diffuse large
  b-cell lymphoma.
\newblock Clin Cancer Res 17: 7704-11.
\bibAnnoteFile{AneuCancer6}

\bibitem{MS59}
Ricke RM, Jeganathan KB, van Deursen JM (2011) Bub1 overexpression induces
  aneuploidy and tumor formation through aurora b kinase hyperactivation.
\newblock J Cell Biol 193: 1049-64.
\bibAnnoteFile{MS59}

\bibitem{MS66}
Sotillo R, Schvartzman JM, Socci ND, Benezra R (2010) Mad2-induced chromosome
  instability leads to lung tumour relapse after oncogene withdrawal.
\newblock Nature 464: 436-40.
\bibAnnoteFile{MS66}

\bibitem{MS67}
Weaver BAA, Bonday ZQ, Putkey FR, Kops GJPL, Silk AD, et~al. (2003)
  Centromere-associated protein-e is essential for the mammalian mitotic
  checkpoint to prevent aneuploidy due to single chromosome loss.
\newblock J Cell Biol 162: 551-63.
\bibAnnoteFile{MS67}

\bibitem{MS68}
Burrell RA, McClelland SE, Endesfelder D, Groth P, Weller MC, et~al. (2013)
  Replication stress links structural and numerical cancer chromosomal
  instability.
\newblock Nature 494: 492-6.
\bibAnnoteFile{MS68}

\bibitem{MS60}
Barra V, Schillaci T, Lentini L, Costa G, Di~Leonardo A (2012) Bypass of cell
  cycle arrest induced by transient dnmt1 post-transcriptional silencing
  triggers aneuploidy in human cells.
\newblock Cell Div 7: 2.
\bibAnnoteFile{MS60}

\bibitem{Burns:2013}
Burns MB, Lackey L, Carpenter MA, Rathore A, Land AM, et~al. (2013) Apobec3b is
  an enzymatic source of mutation in breast cancer.
\newblock Nature 494: 366-70.
\bibAnnoteFile{Burns:2013}

\bibitem{Loeb:2008}
Loeb LA, Bielas JH, Beckman RA (2008) Cancers exhibit a mutator phenotype:
  clinical implications.
\newblock Cancer Res 68: 3551-7; discussion 3557.
\bibAnnoteFile{Loeb:2008}

\bibitem{Loeb:2011}
Loeb LA (2011) Human cancers express mutator phenotypes: origin, consequences
  and targeting.
\newblock Nat Rev Cancer 11: 450-7.
\bibAnnoteFile{Loeb:2011}

\bibitem{McFarland:2013}
McFarland CD, Korolev KS, Kryukov GV, Sunyaev SR, Mirny LA (2013) Impact of
  deleterious passenger mutations on cancer progression.
\newblock Proc Natl Acad Sci U S A 110: 2910-5.
\bibAnnoteFile{McFarland:2013}

\bibitem{Burgess:2013}
Burgess DJ (2013) Tumour evolution: Weighed down by passengers?
\newblock Nat Rev Cancer 13: 219.
\bibAnnoteFile{Burgess:2013}

\bibitem{MS62}
Rutherford S, Hirate Y, Swalla BJ (2007) The hsp90 capacitor, developmental
  remodeling, and evolution: the robustness of gene networks and the curious
  evolvability of metamorphosis.
\newblock Crit Rev Biochem Mol Biol 42: 355-72.
\bibAnnoteFile{MS62}

\bibitem{MS61}
Yeyati PL, van Heyningen V (2008) Incapacitating the evolutionary capacitor:
  Hsp90 modulation of disease.
\newblock Curr Opin Genet Dev 18: 264-72.
\bibAnnoteFile{MS61}

\bibitem{MS63}
Levy SF, Siegal ML (2008) Network hubs buffer environmental variation in
  saccharomyces cerevisiae.
\newblock PLoS Biol 6: e264.
\bibAnnoteFile{MS63}

\bibitem{MS64}
Bester AC, Roniger M, Oren YS, Im MM, Sarni D, et~al. (2011) Nucleotide
  deficiency promotes genomic instability in early stages of cancer
  development.
\newblock Cell 145: 435-46.
\bibAnnoteFile{MS64}

\end{thebibliography}



\end{document}